\documentclass{article}
\usepackage{wrapfig}
\usepackage{amssymb}
\usepackage{amscd,amsmath,amssymb}
\usepackage{graphicx}
\usepackage{amscd}

\def\t{\theta}

\def\D{\Delta}

\def\be{\begin{equation}}
\def\ee{\end{equation}}
\def\arr{\begin{array}{rll}}
\def\ea{\end{array}}
\def\bea{\begin{eqnarray}}
\def\eea{\end{eqnarray}}

\def\N2{$N{=}2$}

\def\>{\rangle}
\def\<{\langle}
\def\+{\dagger}
\def\={\ =\ }

\begin{document}

\setcounter{page}{1}
\begin{center}
{\Large\bf  Action-angle variables for the particle near  extreme Kerr throat }\\

\vspace{0.5 cm} {\large Stefano Bellucci$\;^{a}$, Armen Nersessian$\;^{b}$ and Vahagn Yeghikyan
$\;^{a,b}$ }
\end{center}
$\;^a${\sl INFN-Laboratori Nazionali di Frascati,Via E. Fermi 40,
00044,
Frascati, Italy}\\
$\;^b${\sl Yerevan State University,
1 Alex Manoogian St., Yerevan, 0025, Armenia}
\begin{abstract}
 \noindent
We construct the action-angle variables for the spherical part of
conformal mechanics describing the motion of a particle near
extreme Kerr throat. We indicate the existence of  the critical
point $|p_\varphi|=mc R_{\rm Sch}$ (with $m$ being the mass of the
particle, $c$ denoting the speed of light, $R_{\rm Sch}=2\gamma M
/c^2$ being the Schwarzschild radius of a black hole with mass
$M$, and $\gamma$ denoting the gravitational constant), where
these variables are expressed in terms of elementary functions.
Away from this point the action-angle variables are defined by
elliptic integrals.
 The
proposed formulation allows one to easily reconstruct the whole
dynamics of the particle both in initial coordinates, as well as
in the so-called conformal basis, where the Hamiltonian takes the
form of conventional non-relativistic conformal mechanics.
 The related issues, such as semiclassical quantization and  supersymmetrization  are  also discussed.

\end{abstract}
\numberwithin{equation}{section}

 \section{Introduction}

The Kerr solution \cite{kerr} was discovered in 1963 as a solution of the vacuum Einstein equations describing the rotational black holes.
Its uniqueness, proven  by Carter \cite{car71}, as well as the separability of variables of  the particle moving
 in the Kerr background \cite{car} gave to the Kerr solution a special role in General Relativity.

A very particular case of the Kerr solution, when  the Cauchy's
and event horizons coincide is called extreme Kerr  solution. In
this special case the angular momentum $J$ of the Kerr black hole
is related with the mass of the Kerr black hole $M$ by the
expression $J=\gamma M^2/c$ (with $\gamma$ being the gravitational
constant and $c$ being the velocity of light). In 1999 Bardeen and
Horowitz derived the near-horizon limit of the extreme Kerr
solution and found that the isometry group of the limiting metric
is $SO(1,2)\times U(1)$ \cite{bh}. They conjectured that the
extreme Kerr throat solution might admit a dual conformal theory
description in the spirit of AdS/CFT duality. The extensive study
of AdS/Kerr duality was initiated almost a decade later in
\cite{str} which continues today (see e.g. \cite{cms} and refs
therein). It is clear to the moment, that the near-horizon extreme
Kerr solution and its generalizations
  play a distinguished role in supergravity. Particularly,    their
 thermodynamical properties and connection to the string  theory allow one
 to expect that the quantum gravity should be closely related with these objects.

The simplest  way to study of the near-horizon limit of the extreme Kerr black hole is  the
investigation of  a test particle moving in its field. The study of this system may help to reveal some important symmetries
   or non-trivial constructions related to the field. The direct interpretation of the purely mechanical problem is also motivated,
 since there are known objects with a set of parameters close to those of the extremal Kerr's black hole \cite{astro}.
  As a consequence of extended $SO(1,2)\times U(1)$ symmetry of the near-horizon extreme
Kerr metric, the dynamics  of a test  particle is described by an integrable conformal mechanical model.
 On the other hand, any mechanical system with (dynamical) conformal symmetry can be represented in the ``non-relativistic" form
 \be
 H=\frac{p_{R}^2}{2}+\frac{2{\cal I}(u)}{R^2},
 \ee
 where the role of ``effective coupling constant" ${\cal I}(u)$ is played by the Casimir of the conformal algebra. Here $u$ is a set of phase variables such that $\{u,R\}=\{u,P_R\}=0$. Together with $R$ and $P_R$ the variables $u$ parametrize the entire phase space of the system.
 Hence, the whole specific information on the given  conformal mechanics could be encoded in the ``spherical part"  defined by the
 Hamiltonian ${\cal I}(u)$ and respective Poisson brackets, including (im)possibility to construct the
 $D(1,2|\alpha)$   superconformal extension \cite{hkln}. So, rather than studying the whole dynamics of the probe particle
 near the extreme Kerr throat, we can restrict our study to its spherical part, which defines an integrable
 two-dimensional Hamiltonian system.
Let us notice that the study of the probe particle  near the extreme Kerr throat
 and  of  its ${\cal N}=2$  superconformal  extension has been performed
in \cite{anton}(see also \cite{BelKriv}), and extended in  \cite{anor} to the case of the
 Kerr-Newman-AdS-dS black hole.  However there were troubles  with the
construction of  its ${\cal N}=4$ superconformal extension, which  reflected the lack of supersymmetry invariance of
the extreme  Kerr solution. The investigation of the spherical part  should clarify weather these troubles  are crucial.\\

 The goal of present paper is the construction of action-angle variables for the integrable
 two-dimensional system,  which plays the role of the spherical part of the dynamic of the particle  moving near an extreme Kerr throat.
 The action-angle variables would provide us with the complete information  on  (quasi)periodic motions in  the system.
 Because of the uniqueness among all the other canonical  variables, these variables allow us to establish a correspondence/discrepancy between different integrable systems of classical mechanics, to perform their  Bohr-Sommerfeld quantization, as well as  to construct  their supersymmetric (at least, formally) extensions.
Besides,  action-angle variables forms a tool for the developing of classical perturbation theory.

The construction of the above mentioned variables $u$ is not a straightforward task. In \cite{gn} the construction was done using action-angle variables. Thus, as a by-product of the present studies we present explicit expressions for these variables.

 Performing the action-angle formulation of the spherical part of the ``near-horizon extreme Kerr particle" we will find,
   that the system under consideration has a critical point
   defined by the value of angular momentum $|p_\varphi|$,
   \be
   |p_\varphi|=\gamma \frac{2m M}{c} ,
   \label{p}\ee
   with $m$ being the mass of the particle and  $M$ denoting the black hole mass.
     At this critical point the action-angle variables are given in terms of elementary functions,
     while the Hamiltonian coincides with that of the one-dimensional Higgs oscillator.
     Respectively, the system becomes exactly solvable. Away from the critical point the  functional dependence of the action-angle variables
     from the initial ones, as well as the dependence of the Hamiltonian from the action variables
     are given by elliptic integrals. For the extreme Kerr throat, the existence of this critical point (as the point where the particle
     motion becomes integrable in terms of
 elementary functions) has been noticed already in Ref. \cite{anton}.\\

  The paper is arranged as follows:

  In {\sl Section 2} we present the general consideration and show that the phase space of the spherical part of the particle moving near
  the extreme Kerr throat has a critical point (\ref{p}). In {\sl Section 3} we construct
  the action-angle variables for the system. In {\sl Section 4} we restore the radial dynamics of the probe particle.
  Finally, in {\sl Section 5} we present some concluding remarks and discuss the issues concerning supersymmetrization.

    Throughout the text we will use the gravitational units, where $\gamma = c =1 $(where $\gamma$ is the gravitational constant and $c$ is the
    velocity of light).

\setcounter{equation}{0}
\section{General consideration}

The Kerr solution is the stationary axially symmetric solution of the vacuum Einstein equations,
which describes the rotating black hole with mass $M$  and angular momentum $J$. It is defined by the metric
\be\label{kerr}
{d s}^2=-\left(\frac{\Delta-a^2 \sin^2{\t}}{\Sigma} \right) dt^2
+\frac{\Sigma}{\D} d r^2+\Sigma d \t^2 +
\left(\frac{{(r^2+a^2)}^2-\Delta a^2 \sin^2{\t}}{\Sigma} \right) \sin^2{\t} d \varphi^2
-\frac{2 a  (r^2+a^2-\Delta) \sin^2{\t}}{\Sigma} dt d \varphi,
\ee
where
\be
\D=r^2+a^2-2 Mr, \qquad \Sigma=r^2+a^2 \cos^2 {\t},\qquad a= \frac{J}{M}.
\ee

The extreme solution of the Kerr metric corresponds to the choice  $M^2=J$, so that the event horizon is at  $r=M$.
The limiting near-horizon metric  is given by the expression \cite{bh}
\be\label{kn}
{d s}^2=\left(\frac{1+\cos^2 {\t}}{2} \right) \left[-\frac{r^2}{r_0^2} dt^2+\frac{r_0^2}{r^2} dr^2+r_0^2 d \t^2 \right]+
\frac{2 r_0^2 \sin^2{\t}}{1+\cos^2 {\t} } {\left[d \varphi+\frac{r}{r_0^2} dt \right]}^2,\qquad r_0\equiv \sqrt{2}M.
\ee
The Kerr metric  admits the second rank Killing tensor \cite{wp}, which allows to
integrate the geodesic equation for a massive particle in Kerr space-time by quadratures \cite{car}.
The  limiting
Killing tensor  becomes  reducible, in the sense that
it can be constructed from the Killing vectors corresponding to the $SO(2,1) \times U(1)$ isometry group.

Consequently, the motion of a particle near the horizon of the extreme Kerr black hole
is defined by the generators of   the conformal  algebra $so(1,2)$ \cite{anton}
\be\label{h} H=\frac{r}{r^2_0}\left(\sqrt{{(r p_r)}^2 +L(p_\theta,p_\varphi, \theta)
} -p_\varphi\right),\quad K=\frac{r^2_0}{r} \left(\sqrt{ {(r p_r)}^2
+L(p_\theta,p_\varphi, \theta) } +p_\varphi\right) \quad
D=r p_r, \ee
\be\label{confalg} \{H,D \}=H, \quad \{H,K \}=2D, \quad \{D,K \} =K,
\ee
where Poisson brackets are defined by the canonical symplectic structure
\be
\omega=dr\wedge d p_r +d\varphi\wedge d p_\varphi +d\theta\wedge d p_\theta\;.
\ee
The generator $H$ plays the role of the Hamiltonian  of the system. It has two constants of motion,
\be
L=p_\theta^2+\frac{(1+\cos^2 {\theta})^2 p^2_\varphi}{4\sin^2\theta}  +\left(\frac{1+
\cos^2 {\theta}}{2} \right) {(m r_0)}^2,
\ee
defined by the Killing tensor of the second rank,
and
\be
p_\varphi\; :\{p_\varphi, H\}=0,
\ee
corresponding to the $U(1)$ isometry of the near horizon extreme Kerr metric.\\

The angular part of the system, given by the Casimir of $so(1,2)$
defines the following integrable Hamiltonian system:
\be
{\cal I}= KH-D^2=L-p^2_\varphi=p_\theta^2+
\left(\frac{\cos^4\theta+6\cos^2\theta-3}{4\sin^2\theta}\right) p^2_\varphi +\left(\frac{1+
\cos^2 {\theta}}{2} \right) {(m r_0)}^2,
\qquad \omega_0=dp_\theta\wedge d\theta+dp_\varphi\wedge d\varphi.\label{calint}
\ee
Let us notice, that effective metric of the configuration space of the system
 has the singularities at latitudes defined by angles $\theta_0 =\arccos\sqrt{2\sqrt{3}-3}$ and $\pi-\theta_0$.\footnote{We can also extract the spherical part for the system of a particle moving on the near-extreme Kerr-Newman background and perform similar analyses, however, because of the complexity of the resulting system, we avoid including explicit formulae in this work.}

Besides, the above system  has some distinguished points in the phase space as well,
which could be visualized after formulating the system in terms of action-angle variables.
 In accordance with Liouville's theorem, the existence of action-angle variables requires
 the level surface to be a compact and connective manifold \cite{arnold}. To show that it is a case,
  let us re-write the Hamiltonian of the system in the following form:
\be\label{itx}
{\cal I}=\frac{\left(p_\varphi^2-2(mr_0)^2\right)\cos^4{\theta}+6p_\varphi^2 \cos^2{\theta}+2(mr_0)^2-3p_\varphi^2}{4\sin^2{\theta}}+p_\theta^2.
\ee
From this expression it follows that
${\cal I}\sim{4p_\varphi^2}{/\sin^2{\theta}}>0$ when $\theta\sim0$,
i.e. the requirement of compactness  is satisfied.\\

In order to get the action-angle formulation of the system we have to consider the generating function (``effective action")
\be\label{action}
S(p_\varphi,{\cal I}|, \theta, \varphi)=p_\varphi\varphi+
S_0(p_\varphi,{\cal I},\theta),\ee
where
\be
S_0=
\int_{{\begin{array}{c}{\cal I}=const\\
 p_\varphi=const
 \end{array}}} d \theta\sqrt{{\cal I}- \left(\frac{\cos^4\theta+6\cos^2\theta-3}{4\sin^2\theta}\right) p^2_\varphi
 -  \left(\frac{1+
\cos^2 {\theta}}{2} \right) {(m r_0)}^2} \;.
\ee
By its use  we define  the action variables,
\be
I_1({\cal I}, p_\varphi)=\frac{1}{2\pi}\oint d \theta\sqrt{{\cal I}-
 \left(\frac{\cos^4\theta+6\cos^2\theta-3}{4\sin^2\theta}\right) p^2_\varphi-
\left(\frac{1+\cos^2 {\theta}}{2} \right) {(m r_0)}^2}, \qquad
I_2=p_\varphi , \label{I1}\ee
and the angle  ones
\be
\Phi_1=\frac{\partial S_0(p_\varphi,{\cal I},\theta)}{\partial  I_1}, \qquad
\Phi_2=\varphi+\frac{\partial S_0}{\partial I_2}.\label{ang12} \ee

From \eqref{itx} one can notice that the condition
$\left(p_\varphi^2-2(mr_0)^2\right)=0$  simplifies the expression
for ${\cal I}$ and makes it exactly coincide with the Hamiltonian
of the Higgs oscillator. We will discuss this particular case in
more detail later in this article.

\section{Action-angle variables}

In this Section first we assume that $|p_\varphi |\neq
\sqrt{2}mr_0$. The case $|p_\varphi |= \sqrt{2}mr_0$ will be
analyzed in a separate subsection.

Introducing the variable $x=\cos^2{\theta}$,  we transform \eqref{I1} to the following form:
\be
I_1({\cal I}, p_\varphi)=\frac{1}{2\pi}\oint \frac{d x}{2\sqrt{x(1-x)}}\sqrt{\frac{(2(mr_0)^2-p_\varphi^2)(x-a_1)(x-a_2)}{4(1-x)}},
\label{I1new}
\ee
where
\be
a_{1,2}=\frac{2{\cal I}+3p_\varphi^2\mp2\sqrt{{\cal I}^2-2{\cal I}(mr_0)^2+(mr_0)^4+
4{\cal I}p_\varphi^2-2(mr_0)^2p_\varphi^2+3p_\varphi^4}}{2(mr_0)^2-p_\varphi^2}.
\ee

In order to calculate the action variable $I_1$,  we have to
determine  the integration range in the first integral in
\eqref{I1new}. For this purpose, we notice that for any values of
the parameters in the considered range we have: \be
(2(mr_0)^2-p_\varphi^2)a_2>0,\quad 1>a_1>0 \ee
 and, from the requirement that the subroot expression in \eqref{I1new} is positive (and, therefore,
 the integrand is a real function), we find that the integration range is always $[0,a_1]$.

Since the substitution of variable $x=\cos^2{\theta}$ identifies
the points $\pm \cos{\theta}$, this corresponds to only one
quarter of a cycle. Therefore, the final expression should be
multiplied by $4$. As a result, we get \be
 I_1({\cal I},p_\varphi)=
     \frac{\sqrt{a_2(2(mr_0)^2-p_\varphi^2)}}{4}a_1F_1(1/2,1,-1/2,2,a_1,a_1/a_2)
\label{finalI}
.\ee
%
where $F_1(1/2,1,-1/2,2,a_1,a_1/a_2)$ is the Appel's hypergeometric function (see \cite{prudnikov} and Appendix).

The angle variables are expressed via elliptic integrals  of the first, second and third kind
(see e.g. \cite{prudnikov,dwight} and Appendix):
\be
 \Phi_1({\cal I},p_\varphi,\theta,\varphi)=
     \frac{1}{\sqrt{a_2((2(mr_0)^2)-p_\varphi^2)}}\frac{\partial {\cal I}}{\partial I_1}
     F\left(\arcsin\frac{\cos{\theta}}{\sqrt{a_1}},\frac{a_1}{a_2}\right)
\label{finalfi1}
\ee
and
\be
 \Phi_2({\cal I},p_\varphi,\theta,\varphi)-\varphi =
     \frac{\frac{\partial {\cal I}}{\partial I_2}+
     \frac{I_2}{2}\left(7+a_2\right)}{\sqrt{a_2(2(mr_0)^2-p_\varphi^2)}}F\left(\arcsin\left(\frac{\cos{\theta}}{\sqrt{a_1}}\right);\frac{a_1}{a_2}\right)
     -\frac{I_2}{2}\sqrt{\frac{a_2}{2(mr_0)^2-p_\varphi^2}}E\left(\arcsin\left(\frac{\cos{\theta}}{\sqrt{a_1}}\right);\frac{a_1}{a_2}\right)-
\ee
$$
-\frac{2I_2}{\sqrt{a_2(2(mr_0)^2-p_\varphi^2)}}
     \Pi\left(a_1,\arcsin\left(\frac{\cos{\theta}}{\sqrt{a_1}}\right);\frac{a_1}{a_2}\right),
\label{finalfI2}
$$
The effective frequencies $\partial{{\cal I}}/\partial I_{1,2}$ can be found  from \eqref{finalI}:
\be
\frac{\partial\cal I}{\partial I_1}=\left(\frac{\partial I_1}{\partial\cal I}\right)^{-1},\qquad
\frac{\partial\cal I}{\partial I_2}=-\frac{{\partial I_1}/{\partial p_\varphi}}{{\partial I_1}/{\partial\cal I}}.
\ee
Namely, we have:

\be
\Omega_1=\frac{\partial {\cal I}}{\partial I_1}=\frac{\pi\sqrt{a_2((2(mr_0)^2)-p_\varphi^2)}}{2 K\left(\frac{a_1}{a_2}\right)}\label{omega1}
\ee
and
\be
\Omega_2=\frac{\partial {\cal I}}{\partial I_2}=-\frac{p_\varphi}{2 K\left(\frac{a_1}{a_2}\right)}\left((7+a_2)K\left(\frac{a_1}{a_2}\right)-a_2E\left(\frac{a_1}{a_2}\right)-4\Pi\left(a_1,\frac{a_1}{a_2}\right)\right),\label{omega2}
\ee
where $K(\phi), E(\pi)$ and $\Pi(n,\phi)$ are complete elliptic integrals of the first, second and the third kind respectively.

\subsection*{Critical point}

At the  critical  point $ |p_\varphi|=\sqrt{2}mr_0 $ the spherical
Hamiltonian ${\cal I}$ reduces to the   Hamiltonian of the
one-dimensional  Higgs oscillator \cite{higgs}: \be {\cal
I}_{osc}=p_\theta^2+2(mr_0)^2\cot^2{\theta}-(mr_0)^2. \ee Its
action-angle  variables can be easily  calculated either directly
by putting $|p_\varphi|=\sqrt{2}mr_0$ in \eqref{ang12} (see, e.g.
\cite{hlnsy}) or by  taking the limit
$\lim\limits_{|p_\varphi|\to\sqrt{2}mr_0}I_1$. In both cases we
come to the same result, \be I_1=\sqrt{{\cal
I}_{osc}+3(mr_0)^2}-\sqrt{2}mr_0,\qquad
\Phi_1=\arcsin{\sqrt{\frac{{\cal I}_{osc}+3(mr_0)^2}{{\cal
I}+(mr_0)^2}} \cos{\theta}}. \ee Respectively, the Hamiltonian
reads  \be {\cal
I}_{osc}=\left(I_1+\sqrt{2}mr_0\right)^2-3(mr_0)^2. \ee

This point, however, does not correspond to any special values of
the frequencies. One can see this by taking the limit from
\eqref{omega1},\eqref{omega2}: \be
\Omega_1^{osc}=\lim\limits_{p_\varphi^2\to2(mr_0)^2}\Omega_1=\frac{d
{\cal I}_{osc}}{d I_1}=2\sqrt{{\cal
I}+3(mr_0)^2}=2\left(I_1+\sqrt{2}mr_0\right) , \ee while the
second frequency reads:
$$
\Omega_2^{osc}=\lim\limits_{p_\varphi\to\pm\sqrt{2}mr_0}\Omega_2=\pm\sqrt{2}mr_0\left(\frac{15}{4}-
\frac{(mr_0)^2}{2({\cal I}+3(mr_0)^2)}-\frac{\sqrt{2}}{mr_0}\sqrt{{\cal I}+3(mr_0)^2}\right)=$$
\be
=\pm\sqrt{2}mr_0\left(\frac{15}{4}-
\frac{(mr_0)^2}{4(I_1+\sqrt{2}mr_0)}-\frac{\sqrt{2}}{mr_0}(I_1+\sqrt{2}mr_0)\right).
\ee

\setcounter{equation}{0}
\section{Radial dynamics}
In previous Sections  we have got the complete description of the ``spherical sector" of the ``near-horizon extreme Kerr particle"
in terms of action-angle variables.
 For the restoring of the  radial dynamics of particle, one should  simply replace,  in
 (\ref{h}),  the
$L(p_\varphi, p_\theta, \varphi, \theta)$ by $L(I)={\cal I}(I_{1,2})-I^2_2$, and immediately
solve  the  equations of motion.

However, it is more deductive to write down the whole Hamiltonian systems  in the ``conformal basis", where
the Hamiltonian takes the form of conventional non-relativistic conformal mechanics \cite{gn}
\be
H=\frac{P_{X}^2}{2}+\frac{2{\cal I}(I_{1,2})}{X^2},\qquad
\Omega=dP_X\wedge dX+dI_a\wedge d{\widetilde\Phi}^a,
\ee
where
\be
X=\sqrt{2K(r, p_r, I)},\qquad P_X=\frac{D(r, p_r, I)}{\sqrt{2K(r,p_r, I)}},\qquad
{\widetilde\Phi^a}=\Phi^a+\frac12\frac{\partial}{\partial I_a}\int_{s=p_rr} d s\; {\log\left(\sqrt{s^2/4
+L(I) } +I_2\right) }.
\label{tphi}\ee
The classical properties of this system are well-known \cite{ll}.
Particularly,  for ${\cal I}<0$ we deal with the phenomenon of falling towards the black hole center, and for
 ${\cal I}>0$ we deal with the scattering problem. The case ${\cal I}=0$ corresponds to the free particle system.\\

Quantum mechanics of this system is also considered in details \cite{llq}.
It is well known, that the respective quantum Hamiltonian
is not self-conjugated, and does not possess a ground state \cite{case,fubini}.

For applying this well-known results to our concrete case, we
should simply perform the Bohr-Sommerfeld quantization of the spherical part of the system under consideration
\be
I_1=\hbar(n_1+\frac{1}{2}),\quad  I_2=\hbar n_2,\quad n_1,m_2\in \mathbb{N},
\ee
and consider the quantized ${\cal I}(n_1, n_2)$ as an effective coupling constant.

%

Thus,   upon consideration of the scattering problem, the dependence of the angular Hamiltonian ${\cal I}$ on action
  variables $I_{1,2}$
  will be reflected at the dynamics of radial variables.
 In the noncritical regimes the scattering of
  the particle will be asymmetric (and depending on both quantum numbers $n_1, n_2$),
and at the critical point $n_2 =\sqrt{2}mr_0/\hbar $ we shall deal with the symmetric scattering.
At the critical point we have an exact
coincidence of the spectra obtained by  canonical and Bohr-Sommerfeld quantization.
 The expression for the energy levels cannot be found analytically, but for
  given values of parameters $m r_0$ one can calculate them numerically from \eqref{finalI}.

\section{Concluding remarks}
In the present paper we constructed the action-angle variables of the angular sector of the dynamics
 of a particle moving near an extreme Kerr throat. These variables can be expressed via initial ones in terms of elliptic functions,
so the procedure is not very convenient
for analyzing the system.

Due to the dynamical conformal symmetry, the ``angular part" of
the system accumulates the whole information on   the initial
dynamics. Moreover, it allows us to present the system in the form
of conventional non-relativistic  conformal mechanics, where the
Casimir of the conformal algebra  (``spherical Hamiltonian")
${\cal I}$
 appears as an effective coupling constant.
Respectively, for negative  values of the Casimir, the effective radial dynamics corresponds to the falling on the center,
and for positive Casimir values it corresponds to the scattering problem.

The given formulation allows us to immediately construct the
$D(1,2|\alpha)$ superconformal extension of the particle near the
extreme Kerr throat. Indeed, in action-angle variables we can
construct a (formal) ${\cal N}=4$ superextension of the system
(\eqref{calint}) \be Q_1=\sqrt{{\cal I}}(\cos\lambda e^{\rm \imath
\kappa}\eta_1 -\sin\lambda e^{\rm -\imath \kappa}\eta_2)\; ,\quad
Q_2=\sqrt{{\cal I}} (\cos\lambda e^{\rm -\imath \kappa}\eta_2
+\sin\lambda e^{\rm \imath \kappa}\eta_1) \;: \ee \be
\{Q_a,{\overline Q}_b\}=\delta_{ab}({\cal I}+...),\quad \{Q_a,{
Q}_b\}=0,\quad a,b=1,2 \ee where $\lambda(I,\Phi ) $ and $\kappa
(I,\Phi)$ are some, still undefined  functions, the Grassmann
variable $\eta_a$ obeys the canonical Poisson bracket relations
$\{\eta_a,{\overline\eta}_b\}=\delta_{a,b}$,  and $"..."$ denotes
terms containing fermionic degrees of freedom, which we do not
present explicitly.  By an appropriate choice
   of $\lambda$ and $\kappa$ one can, seemingly, obtain a
 physically relevant supersymmetric Hamiltonian, in analogy with
 the ${\cal N}=2$ case \cite{lny} (here we should exploit, in fact,  the freedom in the supersymmetrization
 of mechanical systems, see  e.g. \cite{freedom,anor}). Then, with the ${\cal N}=4$ supersymmetric extension of the angular Hamiltonian ${\cal I}$
 at hands,  one can
immediately  construct the $D(1,2|\alpha)$ superconformal extension of the whole conformal mechanics \cite{hkln}.

This looks as a contradiction of the customary statement that it
is not possible to construct the ${\cal N}=4$ superconformal
extension of particle systems near the extreme Kerr throat, which
reflects the lack of supersymmetry of the Kerr solution. The
matter is that the transition to the action-angle variables has
been done by a canonical transformation mixing initial coordinates
and momenta. Obviously, the respective  field transformation  is
forbidden for the field-theoretical equations (which admit extreme
Kerr solution). Similarly, {\sl no go} statements on
 ${\cal N}=4 $ (non)supersymmerization of the particle near the extreme Kerr throat (and of the two-dimensional spherical system (\eqref{calint})
 concern the ${\cal N}=4$ superextension
  in terms  of existing one-dimensional  supermultiplets constructed over the initial spatial variables.

   Finally, let us notice that
the  action-angle variables  yield a ground for developing
   classical perturbation theory. Thus, by their use we can   describe  the
    dynamics of a particle in the field of nearly-extreme Kerr black holes,
    which  seemingly, have been observed recently \cite{astro}. Moreover, action variables, being
    adiabatically invariant, allow one to evaluate the particle dynamics near the extreme Kerr throat
    with the slow time-dependent parameters $m$ and $r_0$.\\

{\large Acknowledgements.} We thank Roland Avagyan and Anton
Galajinsky for useful discussions and comments. This work was
supported  by the Volkswagen Foundation grant I/84 496,  the
Armenian State Committee of Science  grants 11-1c258 and SCS-BFBR
11AB-001, and   by the ERC Advanced Grant no. 226455,
Supersymmetry, Quantum Gravity and Gauge Fields (SUPERFIELDS).

\appendix
\setcounter{equation}{0}

\section{Elliptic Integrals}
For the convenience of the reader  we present here the  notions of the elliptic integrals \cite{dwight,prudnikov}
\subsubsection*{ Appel's hypergeometric function}
\be
F_1(\alpha,\rho,\lambda,\alpha+\beta;ua,va)=a^{1-\alpha-\beta}
\frac{\Gamma(\alpha+\beta)}{\Gamma(\alpha)\Gamma(\beta)}
 \int\limits_0^a x^{\alpha-1}(a-x)^{\beta-1}(1-ux)^{-\rho}(1-vx)^{-\lambda}dx .
\nonumber\ee

\subsubsection*{Elliptic integral of the first kind}
\be
F(\varphi, k)=\int_{0}^{\varphi}\frac{d\varphi}{\sqrt{1-k^2\sin^2\varphi}}=\int_{0}^{x}\frac{dx}{\sqrt((1-x^2){1-k^2x^2)}},
\qquad k^2<1, \; x=\sin\varphi
\nonumber\ee
\subsubsection*{Elliptic integral of the second kind}
\be
E(\varphi, k)=\int_{0}^{\varphi}{\sqrt{1-k^2\sin^2\varphi}}{d\varphi}=\int_{0}^{x}\frac{\sqrt{1-k^2x^2}}{\sqrt{1-x^2}}dx ,\qquad  x=\sin\varphi
\nonumber\ee
\subsubsection*{Elliptic integral of the third kind}
\be
\Pi (\varphi,n, k)=\int_{0}^{\varphi}\frac{d\varphi}{(1+n\sin^2\varphi){\sqrt{1-k^2\sin^2\varphi}}}=
\int_{0}^{x}\frac{dx}{(1+nx^2)\sqrt{1-x^2}\sqrt{1-k^2x^2}}dx ,\qquad  x=\sin\varphi
\nonumber\ee
\subsubsection*{Complete elliptic integral}
\be
K\equiv F(\pi/2, k)=\int_{0}^{\pi/2}\frac{d\varphi}{\sqrt{1-k^2\sin^2\varphi}}.
\nonumber\ee
The detailed description of elliptic integrals can be found, e.g. in \cite{dwight,prudnikov}.


\begin{thebibliography}{99}
\bibitem{kerr}
  R.~P.~Kerr,
  Phys.\ Rev.\ Lett.\  {\bf 11} (1963) 237.

\bibitem{car71} B.~Carter, Phys.\ Rev.\ Lett.\  {\bf 26}, 331 (1971).
\bibitem{car} B.~Carter, Phys. Rev. {\bf 174}, 1559 (1968).
\bibitem{bh} J.~M.~Bardeen, G.T. Horowitz, Phys. Rev. D {\bf 60},  104030 (1999) [hep-th/9905099].
\bibitem{str}
M. Guica, T. Hartman, W. Song, A. Strominger, Phys.\ Rev.\ D\  {\bf 80}, 124008 (2009)
[arXiv:0809.4266].

\bibitem{cms}
A.~Castro, A.~Maloney, A.~Strominger, Phys.\ Rev.\ D \ {\bf 82},  024008 (2010)
[arXiv:1004.0996].

\bibitem{astro}
J.~E.~McClintock, R.~Shafee, R.~Narayan, R.~A.~Remillard, S.~W.~Davis, Li-Xin~Li,
Astrophys.\ J.\ {\bf 652}, 518 (2006)[astro-ph/0606076]

\bibitem{hkln}
T. Hakobyan, S. Krivonos, O. Lechtenfeld, A. Nersessian, Phys.\
Lett. A {\bf 374} (2010) 801 [arXiv:0908.3290];


\bibitem{anton}A.~Galajinsky,
  JHEP {\bf 1011}, 126 (2010)
  [arXiv:1009.2341].
\bibitem{anor}
 A.~Galajinsky and K.~Orekhov,
  Nucl.\ Phys.\  B {\bf 850}, 339 (2011)
  [arXiv:1103.1047 [hep-th]].
\bibitem{BelKriv}
  S.~Bellucci and S.~Krivonos,
  JHEP {\bf 1110}, 014 (2011)
  [arXiv:1106.4453 [hep-th]].


  \bibitem{wp} M.~Walker, R.~Penrose, Commun. Math. Phys. {\bf 18}, 265 (1970).
\bibitem{arnold}
V.~I.~Arnold, {\sl Mathematical methods in classical mechanics\/},
Nauka Publ., 1973.

 \bibitem{prudnikov}
A.~P.~Prudnikov, Yu.~A.~Brychkov, O.~I.~Mar'ichev, {\sl Integrals and series} (in Russian), Nauka Publ., Moscow, 1981
\bibitem{dwight}H.~B.~Dwight, {\sl Tables of integrals and other mathematical data.}, 4th edition, The Macmillan Company, N.Y., 1961
\bibitem{higgs}P.~W.~Higgs,
 J.\ Phys.\  A {\bf 12}, 309(1979)

H.~I.~Leemon,
 J.\ Phys. \ A {\bf 12}, 489 (1979).
\bibitem{hlnsy}
  T.~Hakobyan, O.~Lechtenfeld, A.~Nersessian, A.~Saghatelian, V.~Yeghikyan,
  Phys.\ Lett.\  {\bf A376} 679 (2012) [arXiv:1108.5189][math-ph]

    S.~Bellucci, A.~Nersessian, A.~Saghatelian, V.~Yeghikyan,
  J.\ Comput.\ Theor.\ Nanosci.\ {\bf 8}, 769 (2011)
[arXiv:1008.3865].

\bibitem{gn} A.~Galajinsky and A.~Nersessian,
 JHEP\ {\bf 1111}, 135 (2011)  [arXiv:1108.3394][hep-th]
\bibitem{ll}L.~D.~Landau, E.~M.~Lifshits, {\sl Mechanics.} 5th edition, Nauka Publ., Moscow, 2004
\bibitem{llq}L.~D.~Landau, E.~M.~.Lifshits,{\sl  Quantum Mechanics.
Non-relativistic theory.}, 4-th edition,
 Nauka Publ., Moscow, 1989
 \bibitem{case}K.~M.~Case, Phys.\ Rev., {\bf 80} (1950) 797
\bibitem{fubini}
 V.~de Alfaro, S.~Fubini and G.~Furlan,
   Nuovo\  Cim. {\bf A34}(1976) 569.


\bibitem{lny}O.~Lechtenfeld, A.~Nersessian, V.~Yeghikyan
Phys.\ Lett.\ A \ {\bf 374} (2010) 4647 [arXiv:1005.0464]


\bibitem{freedom}S.~Bellucci,A.~Nersessian,  Phys.\ Rev.\ D {\bf 73} (2006) 107701
[arXiv:hep-th/0512165].


\end{thebibliography}
\end{document}